\documentclass[pra,superscriptaddress,twocolumn,superscriptaddress,amsmath]{revtex4}

\usepackage{graphicx}% Include figure files
\usepackage{dcolumn}% Align table columns on decimal point
\usepackage{bm}% bold math
\usepackage{physics}
\usepackage{comment}
\usepackage[dvipsnames]{xcolor}

\usepackage{dsfont}
\usepackage{xcolor}
\usepackage{todonotes}
\usepackage{listings}
\usepackage[normalem]{ulem}

\begin{document}
\title{Power-consumption Backdoor in Quantum Key Distribution}
%\date{\today}

\author{Beatriz Lopes da Costa}
\affiliation{Instituto Superior Técnico, Universidade de Lisboa}
\affiliation{Physics of Information and Quantum Technologies Group, Centro de Física e Engenharia de Materiais Avançados (CeFEMA)}\affiliation{PQI — Portuguese Quantum Institute}

\author{Matías R. Bolaños}
\affiliation{Dipartimento di Ingegneria dell’Informazione, Università degli Studi di Padova}

\author{Ricardo Chaves}
\affiliation{Instituto Superior Técnico, Universidade de Lisboa}
\affiliation{Instituto de Engenharia de Sistemas e
Computadores – Investigação e Desenvolvimento (INESC-ID)}

\author{Claudio Narduzzi}
\affiliation{Dipartimento di Ingegneria dell’Informazione, Università degli Studi di Padova}

\author{Marco Avesani}
\affiliation{Dipartimento di Ingegneria dell’Informazione, Università degli Studi di Padova}
\affiliation{Padua Quantum Technologies Research Center, Università degli Studi di Padova}

\author{Davide Giacomo Marangon}
\affiliation{Dipartimento di Ingegneria dell’Informazione, Università degli Studi di Padova}
\affiliation{Padua Quantum Technologies Research Center, Università degli Studi di Padova}

\author{Andrea Stanco}
\affiliation{Dipartimento di Ingegneria dell’Informazione, Università degli Studi di Padova}
\affiliation{Padua Quantum Technologies Research Center, Università degli Studi di Padova}

\author{Giuseppe Vallone}
\affiliation{Dipartimento di Ingegneria dell’Informazione, Università degli Studi di Padova}
\affiliation{Padua Quantum Technologies Research Center, Università degli Studi di Padova}

\author{Paolo Villoresi}
\affiliation{Dipartimento di Ingegneria dell’Informazione, Università degli Studi di Padova}
\affiliation{Padua Quantum Technologies Research Center, Università degli Studi di Padova}

\author{Yasser Omar}
\affiliation{Instituto Superior Técnico, Universidade de Lisboa}
\affiliation{Physics of Information and Quantum Technologies Group, Centro de Física e Engenharia de Materiais Avançados (CeFEMA)}\affiliation{PQI — Portuguese Quantum Institute}

\begin{abstract}
Over the last decades, Quantum Key Distribution (QKD) has risen as a promising solution for secure communications. However, like all cryptographic protocols, QKD implementations can open security vulnerabilities. Until now, the study of physical vulnerabilities in QKD setups has primarily focused on the optical channel. 
In classical cryptoanalysis, power and electromagnetic side-channel analysis are powerful techniques used to access unwanted information about the encryption key in symmetric-key algorithms. In QKD they have rarely been used, since they require an eavesdropper to have access to Alice or Bob's setups. However, security proofs of QKD protocols generally assume that these setups are secure, making it crucial to understand the necessary security measures to ensure this protection. In this work, we propose and implement a power side-channel analysis to a QKD system, by exploiting the power consumption of the electronic driver controlling the electro-optical components of the QKD transmitter. QKD modules typically require very precise electronic drivers, such as Field Programmable Gate Arrays (FPGAs). Here, we show that the FPGA's power consumption can leak information about the QKD operation, and consequently the transmitted key. The analysis was performed on the QKD transmitter at the University of Padua. Our results are consistent and show critical information leakage, having reached a maximum accuracy of $73.35\%$ in predicting transmitted qubits at a $100$ MHz repetition frequency.
\end{abstract}
\maketitle

%--------------------------------------------------------------------------------------------------------INTRODUCTION-----------------------------------------------------------------------------------------------------------%

\section{Introduction} \label{sec:introduction}
Quantum Key Distribution (QKD) leverages the features of quantum physics to offer a highly-secure way for two authenticated parties to share a secret key. From satellite links reaching intercontinental distances \cite{SateliteLinkJeiWeinPan,SateliteLink1200} to fiber networks spanning metropolitan areas \cite{Peev2009,Dynes2019,TokyoNetwork}, QKD is maturing as a commercial quantum technology. Alongside post-quantum cryptography \cite{Bernstein2017}, it offers a promising way to develop cryptographic systems that are secure against attacks by classical and quantum computers \cite{Shor1994AlgorithmsFactoring}. 
\par
However, like any other cryptographic protocol, implementations of QKD systems may contain physical vulnerabilities that can be exploited to obtain key information. The concept of hacking a cryptographic setting via its physical flaws, known as side-channel analysis, was initially introduced in classical cryptanalysis \cite{Kocher1999DifferentialAnalysis}. Classical cryptographic devices are typically implemented using semiconductor transistors as logic gates, where electrons flow across the silicon substrate when voltage is applied or removed from the gate. Consequently, they typically present power consumption and emit electromagnetic radiation that is data-dependent. Exploiting these dependencies has, since then, evolved into the fields of power and electromagnetic side-channel analysis, respectively \cite{Standaert2010SideChannel}.
\par
Conversely, a QKD setting typically involves optical and electronic components. Therefore, physical vulnerabilities in QKD setups may lie in the optical or the electronic band. Using weak coherent state sources as an approximation to single photons or employing non-ideal single photon detectors are examples of physical flaws which have lead to side-channel attacks in the optical band \cite{Huang2018QuantumStates,Makarov2006EffectsCryptosystems}. Regarding the electronic components in Alice's or Bob's setups, it is well known that they must be protected via classical means to guarantee the security of the entire system \cite{Gisin2006Trojan-horseSystems}. In reality, security proofs typically assume that Alice's and Bob's laboratories are secure locations where no information is leaked \cite{Wolf2021QuantumDistribution,Zapatero2023AdvancesDistribution}. Understanding which security measures must be employed to ensure that this assumption holds in practical scenarios is crucial for the safe deployment of QKD systems. To achieve this, the physical vulnerabilities of the electronic components must be characterized. Devices such as Field Programmable Gate Arrays (FPGAs) are ubiquitous to QKD setups \cite{Stanco2022VersatileSystems, Avesani2022Deployment-ReadyPadua, Wei2020High-Speed_QKD, Zhang2012_Real-Time_QKD} and are also based on transistor logic, thus opening the possibility for electronic side-channels. Until now, to our knowledge, work devising these types of side channels for QKD has focused on the electromagnetic band \cite{Baliuka2023Deep-learning-basedDistribution,Durak2022AttackDetectors,Pantoja2024}. Additionally, one should note that electromagnetic side-channel attacks have also been devised, studied and implemented for quantum random number generators (QRNGs) \cite{PhysRevApplied.15.044044,SideChannelQRNG,ExternalMagneticFieldsQRNG}, which are crucial for the security of QKD.  

Regarding power side-channels, using power consumption traces of cryptographic devices to access unwanted information about the encryption key is a standard and powerful technique in cryptoanalysis of classical devices implementing symmetric-key algorithms, such as the Data Encryption Standard (DES) \cite{DESPowerSideChannel} and the Advanced Encryption Standard (AES) \cite{OFlynn2015SideBootloader}. In this work, we aim to devise and implement a power side-channel attack to a component of a QKD transmitter. The side-channel analysis in this work was specifically performed on the System-on-a-Chip (SoC) controlling the electro-optical components of the QKD transmitter at the University of Padua. Since side-channel vulnerabilities are highly implementation-dependent, the findings presented here are specific to the QKD transmitter used.
\par
In this paper, we will start with Sec. \ref{sec:WorkingPrincipleQKDTransmitter} by briefly covering the basic operation of the QKD Transmitter at the University of Padua, focusing on the electronic components. This background is essential for comprehending the experimental setup for the power analysis, detailed in Sec. \ref{sec:Experimental Setup}. Additionally, an understanding of the working principle of the QKD transmitter is presented in this Section to shed light on the potential causes of information leakages. The considered analysis of the SoC's power consumption is divided into two parts: its average value, covered in Sec. \ref{sec:Average Power Consumption}, and its frequency spectrum, covered in Sec. \ref{sec:FrequencySpectrum}. Vulnerabilities found in the analysis of the latter are then used to propose and implement a side-channel attack to the SoC. The methodology of this attack and the results obtained from its implementation are detailed in Sec. \ref{sec:Discovering the Transmitted Qubits}. During the entire analysis, the SoC was treated as a pseudo black-box, as it would be by a current eavesdropper with malicious intent, who would expectedly only have access to publicly available information about the device. Therefore, the only knowledge we assume on the working principle of the SoC is the one made public in \cite{Stanco2022VersatileSystems}, covered in Sec. \ref{sec:WorkingPrincipleQKDTransmitter}. We conclude in Sec. \ref{sec:Conclusions} by evaluating the overall performance of the side-channel attack and the limitations of this approach.

%---------------------------------------------------------------------------------------------Working Principle of the QKD Transmitter-----------------------------------------------------------------------------------------------------------%
\par
\begin{figure*}
    \centering
    \includegraphics[width=0.7\linewidth]{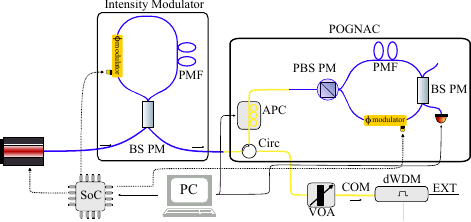}
    \caption{Schematic representation of the QKD Transmitter at the University of Padua}
    \label{fig:QKDTransmitter}
\end{figure*}

\section{Working Principle of the QKD Transmitter} \label{sec:WorkingPrincipleQKDTransmitter}
The QKD setup in Padua implements the three-state one-decoy BB84 protocol with polarization encoding \cite{Grunenfelder2018SimpleQKD}. In the transmitter, an Intensity Modulator (IM) \cite{Roberts2018Patterning-effectDistribution} is used to create decoy and signal states, followed by a polarization encoder, the POGNAC \cite{Agnesi2019All-fiberDistribution}, which encodes the polarization of each photon (Fig. \ref{fig:QKDTransmitter}) \cite{Avesani2022Deployment-ReadyPadua}. 
\par
The three-state BB84 protocol relies on $3$ qubit states, two which contribute to key generation, e.g. $\{\ket{L},\ket{R}\}$, and a third which is solely used for parameter estimation, e.g $\ket{D}$. During QKD operation, the transmitter sends these qubit states to the receiver through the quantum channel at a specific rate known as the qubit repetition frequency $f_{\text{rep}}$. A gain-switched distributed feedback laser is employed as a weak coherent pulse source.
\par
To employ the decoy aspect of the protocol, the intensity of each weak coherent pulse must be modulated between two values, $\mu_1$ and $\mu_2$, one for signal states and the other for decoy states. To achieve this, a Lithium Niobate ($\text{LiNbO}_3$) phase modulator placed inside a Sagnac interferometer is used as a two-level IM \cite{Roberts2018Patterning-effectDistribution}. In this configuration, voltage pulses are sent to the phase modulator at a rate $f_{rep}$, with a voltage chosen between two values, e.g $0$ and $V$, corresponding to the higher and lower intensity respectively. 
\par
After exiting the IM, the pulses have to be encoded with key information. To do so, they are sent through the POGNAC, a polarization encoder based on a $\text{LiNbO}_3$ phase modulator inside a Sagnac interferometer \cite{Agnesi2019All-fiberDistribution}. Here, the light pulses enter the Sagnac loop in the superposition state
\begin{equation}
    \ket{\psi} = \frac{1}{\sqrt{2}}(\ket{H}+\ket{V}),
    \label{eq:POGNAC_InitialPolState}
\end{equation}
where they pass through a Polarizing Beam Splitter (PBS), ensuring that each light pulse is divided into two pulses which travel in opposite directions within the loop. To avoid polarization mode dispersion and birefringence effects, after the PBS, the polarization of one of the pulses is rotated into its orthogonal counterpart, so that only one polarization travels through the fiber. Finally, they arrive to the phase modulator with a time difference of 
\begin{equation}
    \delta t = \frac{\Delta L}{n_fc},
    \label{eq:POGNAC_TimeDifference}
\end{equation}
where $\Delta L$ and $n_f$ are, respectively, the length and the refractive index of a polarization maintaining fiber delay line, placed inside of the loop. In this way, by sending synchronized voltage pulses at $f_{rep}$ and then choosing one of the two time slots separated by $\delta t$, one is able to control the phase applied to each polarization state. By sending an electrical pulse with voltage $V_{\phi}$ \footnote{$V_{\phi}$ corresponds to the voltage required for the modulator to perform a phase shift of $\phi$.} targeting only one of the two polarized pulses (for example, the vertically polarized one), the state $\ket{\psi}=\ket{H}+e^{i\phi}\ket{V}$ can be created. Thus, by choosing $\phi=0,\pm\pi/2$, the states $\ket{D}$, $\ket{R}$ and $\ket{L}$ can be created respectively, which is sufficient for the three-states BB84.
\par 
Overall, the optical encoder in the transmitter requires three separate electrical signals: one to gate the laser, and two for the phase modulators, one in the IM and the other in the POGNAC. To create these electrical pulses with the high temporal resolution required by fast repetition rates $f_{\text{rep}}$, a System-on-a-Chip (SoC) is used \cite{Stanco2022VersatileSystems}.
\par
The SoC comprises a Field Programmable Gate Array (FPGA) and a dual core CPU unit. Its architecture follows a top-down configuration, where data flows from the user/pc to the quantum system, passing through two different layers. Firstly, it goes through the CPU layer, which is responsible for communication with the outside world and data management operations. Then it reaches the FPGA, where all deterministic and high-resolution operations are carried. Finally, from the FPGA, it goes to the chip input-output pins, where the electric pulses are emitted. This is the device whose power consumption we will exploit in a power side-channel attack. 

%-------------------------------------------------------------------------------------------------Experimental Setup-----------------------------------------------------------------------------------------------------------%

\section{Experimental Setup} \label{sec:Experimental Setup}
Our target device is the Zynq-7020 SoC mounted on a ZedBoard by Avnet. The ZedBoard receives a $12$ V input power supply and comes with an inbuilt $10$ m$\Omega$, $1$ W in series resistor for current measurements, placed at the high voltage side of the circuit. This allows us to determine the current drawn from the SoC's power supply and consequently infer its power consumption.
To guarantee enough resolution for measurements of the average power consumption values, the $10$ m$\Omega$ resistance was replaced by an $1~\Omega$, 1~W one. The $1~\Omega$ value was chosen to ensure that voltage across the resistor is large enough to allow resolution in the mV range, while avoiding the need for a higher voltage power supply ($V>12$ V) to the Zedboard.
\par
The voltage difference between terminals of the 1 $\Omega$ resistor translates into the power consumption of the SoC. To characterize it, we acquire the voltage difference during a certain time interval and refer to the resulting data as a power consumption trace.
\par
\begin{figure}[ht]
    \centering
    \includegraphics[width=1\linewidth]{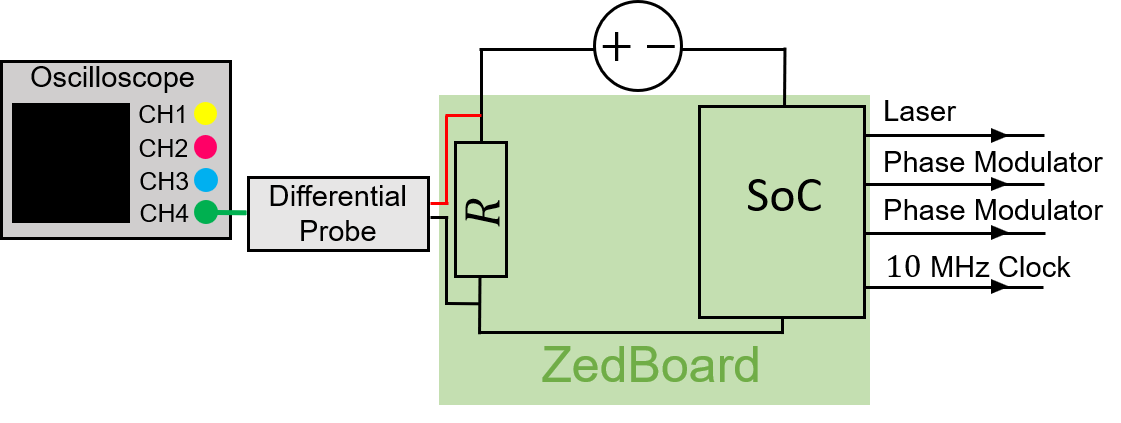}
    \caption{Schematics of the experimental setup with a differential probe.}
    \label{fig:expsetup_differentialprobe}
\end{figure}

The power traces were acquired using the SIGLENT SDS5104X Oscilloscope. Two different probe setups were used to measure the voltage, according to the different requirements of the measurements. The first probe used was a KEYSIGHT N2791A Differential Probe, which directly computes the voltage difference between the two resistance terminals, as schematized in Fig. \ref{fig:expsetup_differentialprobe}. 
This probe has a $25$ MHz bandwidth, which is insufficient for an analysis of the frequency spectrum of the power traces, leaving this setup for the study of the average power consumption of the SoC. 
\par
\begin{figure}[ht]
    \centering
    \includegraphics[width=1\linewidth]{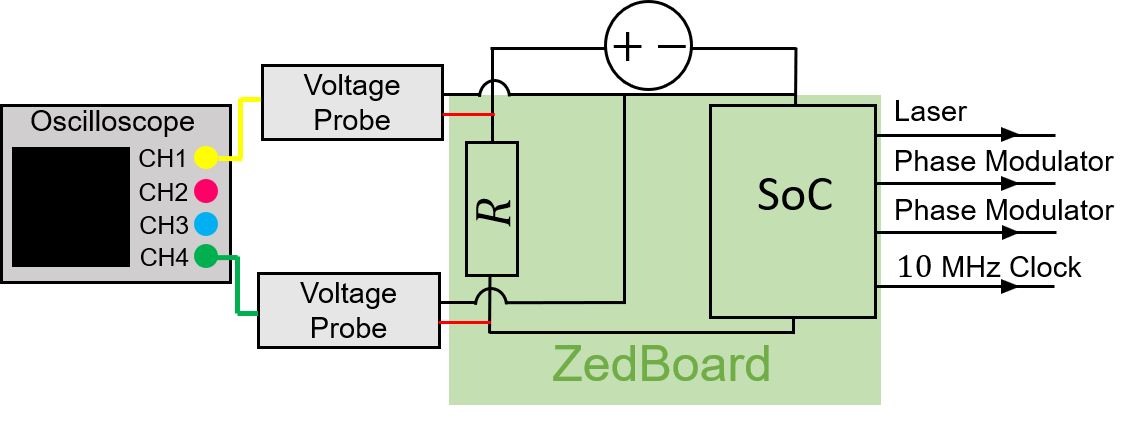}
    \caption{Schematics of the experimental setup with two voltage probes.}
    \label{fig:expsetup_osciprobes}
\end{figure}

For the frequency spectrum analysis of the power traces, two SIGLENT SP$2035$A voltage probes with a bandwidth of $350$ MHz were employed in a pseudo-differential set-up, as illustrated in Fig. \ref{fig:expsetup_osciprobes}. Using the oscilloscope, the power traces were retrieved by computing the difference between the two voltage traces, each measured by one of the probes. In this case, the sampling frequency of the oscilloscope was set to $2.5$ GSa/s, the maximum allowed.
\par
The digital pulse output by the SoC to gain switch the laser was used to trigger the acquisitions in both experimental setups. In addition, to eliminate clock drifts between the SoC and the oscilloscope, a 10 MHz clock generated by the SoC was employed as an external clock source for the oscilloscope
\subsection{System Settings}
To control the system settings we relied on an application which allows the user to set the type of sequence to be sent by directly loading it onto the FPGA in the SoC. For all the measurements taken in this work, the qubit repetition frequency was set to $100$ MHz. Regarding the key, the symbol encoded by each qubit can be chosen between H, V or D, which are interface notations for the states $\ket{L}$, $\ket{R}$ and $\ket{D}$ in Sec.\ref{sec:WorkingPrincipleQKDTransmitter}, respectively. For the purpose of this work, only non-decoy key symbols, H and V, were considered.

In the course of the analysis, two different key types were used. The first one corresponded to fixed sequences, which consist of an infinite repetition of symbols. For example, an Only-H sequence corresponds to having the QKD transmitter sending H symbols indefinitely. Naturally, these sequences would not be used in a practical QKD setting, since to assure theoretical security, each symbol Alice sends must be chosen in a truly random manner, i.e. using a QRNG. Nevertheless, as it will be seen in Sec. \ref{sec:Average Power Consumption}, they are useful to study effects in the SoC's power consumption which manifest at a time interval larger than the emission period of a single symbol $1/f_{\text{rep}}$.
\par
The second type of key is a finite-size random key, which can be used to simulate a more realistic QKD scenario. We note that the random key strings were generated with Python's pseudorandom number generator, thus not corresponding to true randomness. Nevertheless, pseudorandomness provides the framework necessary for validating a possible side-channel attack; thus, from this point forward, we will neglect its difference from true randomness.

%-------------------------------------------------------------------------------------------------Average Power Consumption-----------------------------------------------------------------------------------------------%
\section{Average Power Consumption} \label{sec:Average Power Consumption}
In this section we analyze whether the average power consumption of the FPGA depends on the values of the symbols composing the key. If upon switching between two symbols, i.e H and V, there is a change in the current drawn by the FPGA, this will affect the voltage regulation module in the power supply. The feedback control within this module will ensure that the power supply's voltage remains steady, while the transition between the two current intensity levels occurs smoothly. To calculate the rising time $t_r$, we modeled the behavior of the current during a transient load with

\begin{equation}
    V(t) = \frac{\Delta V}{2}\Big[1+\text{erf}\Big(\frac{\sigma(t-t_0)}{2}\Big)\Big] + V_{t\rightarrow-\infty},
    \label{eq:rise_time_func}
\end{equation}
where $t_r$, defined as the time it takes to go from $10\%$ to $90\%$ of rise, is given by 
\begin{equation}
    t_r = \frac{4}{\sigma}\text{erf}^{-1}(0.8).
    \label{eq:rise_time}
\end{equation}
The rise time was found to be
\begin{equation}
    t_r = (28.64 \pm 0.78) \mspace{5mu} \mu\text{s},
\end{equation}
which is much larger than the emission period of each symbol, i.e. $t_{\text{r}} \gg 10$ ns. Therefore a power consumption dependence on symbol value will not manifest itself in the power trace by an oscillation at $100$ MHz. In other words, at the repetition rate of $100$ MHz, a Simple Power Analysis (SPA) \cite{Kocher1999DifferentialAnalysis} of the power traces will not yield information about the key. 

%---------------------------------------------------------------------------------------------------DEPENDENCE ON SYMBOL VALUE--------------------------------------------------------------------------------------------------------%

\subsection{Dependence on Symbol Value}
To characterize the dependence of the SoC's average power consumption on the symbols composing the key, we considered the emission of fixed sequences with different percentage of H symbols in the key, using a randomized emission method. The fixed sequences considered were Only-H, Only-HHV, Only-HV, Only-HVV and Only-V, chosen given their different percentage of H symbols, respectively $100\%$, $66\%$, $50\%$, $33\%$ and $0\%$.

\par
\begin{figure}[ht]
    \centering
    \includegraphics[width=0.9\linewidth]{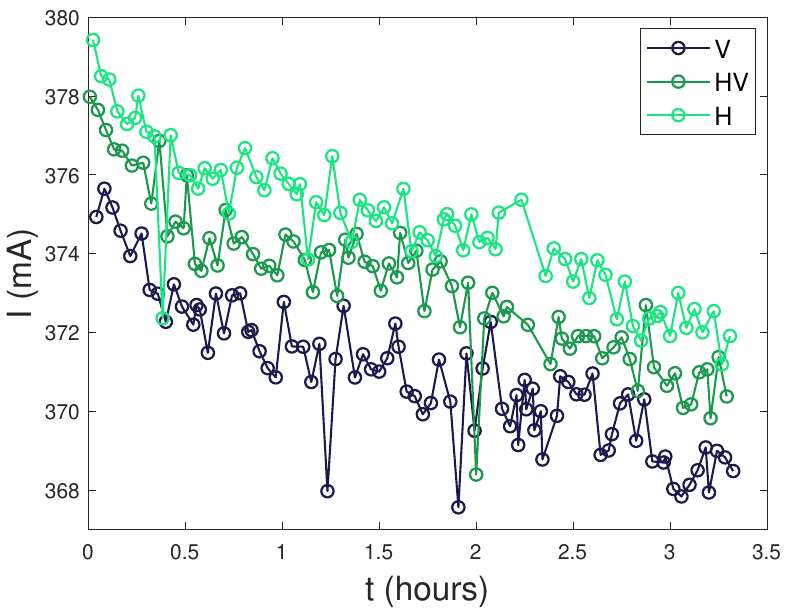}
    \caption{Time evolution of the average power consumption of traces corresponding to different fixed sequences.}
    \label{fig:Time_Evol}
\end{figure}
Regarding the acquisition, the time scale of the oscilloscope was set to $500$ $\mu$s/div, allowing to calculate the average power consumption during the emission of $500000$ symbols with each power trace acquisition. We note that during acquisition, some power traces showed an abnormally low average power consumption. This appeared to affect the traces randomly, which we attributed to faults in the connection between the jack of the power cable and the ZedBoard's socket 
\par
Fig. \ref{fig:Time_Evol} displays the evolution in time of the average power consumption for Only-V, Only-HV and Only-H sequences, where the power consumption drops are visible. The average power consumption for the emissions of each fixed sequence showed a slow decrease over the acquisition time, which may be explained by a temperature dependence of the $1$ $\Omega$ resistor and an overall increase of the resistor's temperature and thus resistance during the measurement phase.
To understand whether this decrease affected all sequences equally, we performed five linear regressions $V = \beta_0 + \beta_1 t$, one for each sequence's time evolution of the average power consumption. From these results, we concluded that the power consumption descent was sequence-independent. 
In order to mitigate its effect, we considered the average value $\overline{\beta_1} = (-1.73 \pm 0.2)$ mA/h, and shifted all data points $(t,V)$ from all sequences to $(t,V-\overline{\beta_1}t)$
\par
\begin{figure}[ht]
    \centering
    \includegraphics[width=0.8\linewidth]{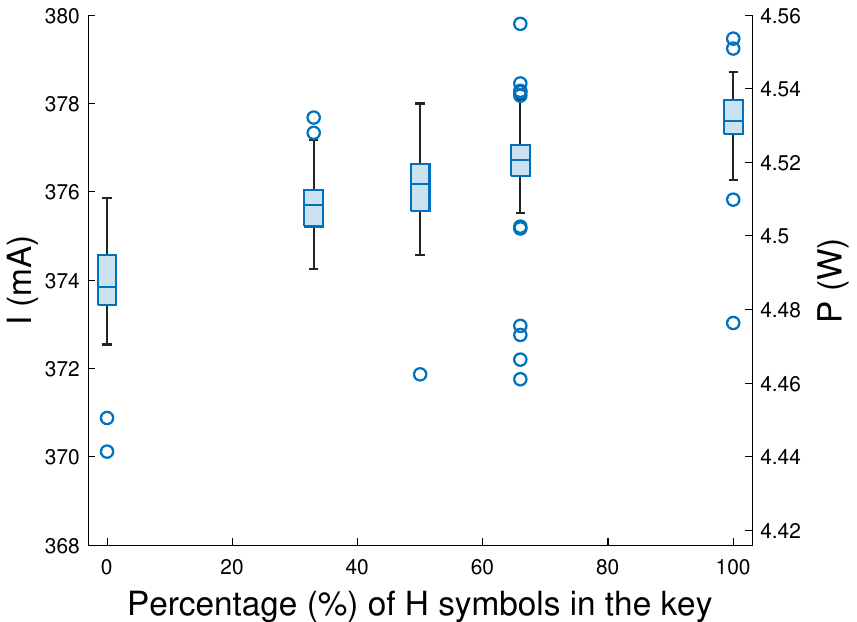}
    \caption{Box plot displaying the average power consumption values for each H symbol percentage. Each data set is represented by its median (the line inside the box), upper and lower quartiles (the top and bottom edges of the box), and whiskers indicating the range of non-outlier data. Outliers are shown as circular points.}
    \label{fig:DCScatter_HSymbPercentage}
\end{figure}

Fig. \ref{fig:DCScatter_HSymbPercentage} shows the average power consumption values for each fixed sequence after mitigating the global power consumption decrease. The current intensity distribution across the $1$ $\Omega$ resistor is displayed for the different percentages of H symbols in the key, together with the corresponding power consumption.
These results showed a trend: the average power consumption increased with the increasing percentage of H symbols in the key. 

%------------------------------------------------------------------------------------------------SIMULATING LOWER REPETITION FREQUENCIES----------------------------------------------------------------------------------------------%
\subsection{Simulating a QKD Transmitter with Lower Qubit Repetition Frequencies}
As aforementioned, given our current acquisition setup, the average power consumption difference between the emission of an H and a V cannot be detected symbol by symbol at a qubit repetition frequency of $100$ MHz. However, it might be detected at lower repetition frequencies. 
\par
It is possible to use the FPGA at $f_{\text{rep}} = 100$ MHz to approximately simulate a smaller qubit repetition frequency $f_{\text{delay}}$. To do so, one emits sequences consisting of $n = f_{\text{rep}}/f_{\text{delay}}$ same valued symbols, each sequence corresponding to the emission of a single symbol at the lower frequency case.
\par
When looking at the distinguishability between H and V emission via the average power consumption, the maximum qubit repetition frequency for which distinguishability is expectedly present corresponds approximately to the bandwidth of our system BW, which we estimate to be
\begin{equation}
    \text{BW} \approx \frac{0.35}{t_\text{r}} = (12.22 \pm 0.95) \mspace{5mu} \text{kHz}.
    \label{eq:power_system_bandwidth}
\end{equation}
All integer qubit repetition frequencies in the range $[2,46]$ kHz were studied. For each repetition frequency, we analysed the power consumption during the emission of a key which simulated the emission of an H followed by a V and so forth at the given lower frequency. 
\par
The time scale of the oscilloscope was set at $500$ $\mu \text{s/div}$, and the delay set to $-2$ ms, meaning that each power trace encoded the rise in power consumption in the first $0.5$ ms and then the emission during $4.5$ ms. The number of key symbols encoded in these $4.5$ ms depends on the repetition frequency we wish to simulate.
\par
\begin{figure}[ht]
    \centering
    \includegraphics[width=0.8\linewidth]{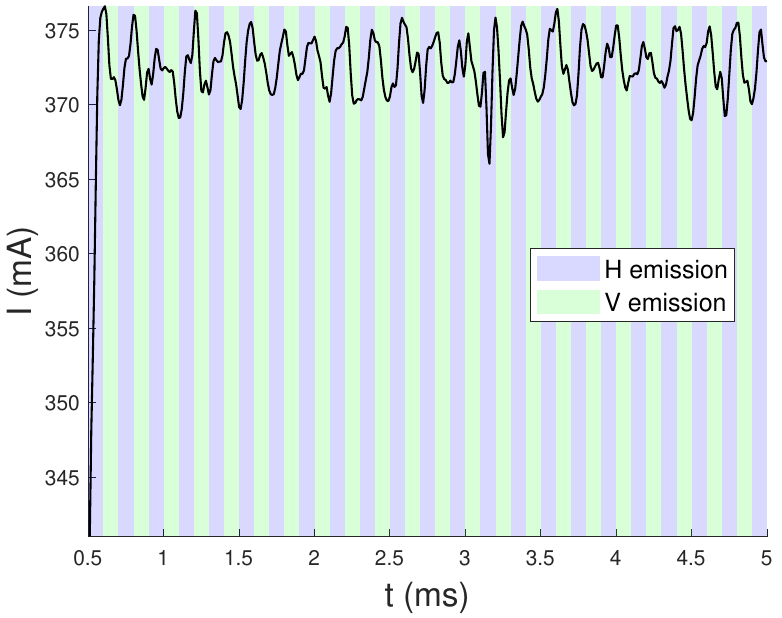}
    \caption{Power trace acquired at $10$ kHz repetition frequency, after filtering and running average treatment.}
    \label{fig:10kPowerTrace}
\end{figure}

In order to reduce the noise, each trace was filtered with a low pass filter, with bandwidth set to the corresponding smaller repetition frequency. Additionally, the AC power supply used so far was changed to a DC power supply to reduce the frequency noise around $20$ kHz introduced by the former. A running average was also performed to all traces, and, to assure coherence between all the frequencies, the number of samples $N_{\text{samp}}$ per average was set to,
\begin{equation}
    N_{\text{samp}} = 0.1 \times \frac{f_{\text{samp}}}{f_{\text{delay}}},
\end{equation}
Fig. \ref{fig:10kPowerTrace} shows an example of a trace acquired at $10$ kHz, after filtering and applying the running average.
In order to computationally distinguish which part of a given treated trace encoded a V and which encoded an H, the power consumption value at the beginning and at the end of each emission period, $1/f_{\text{delay}}$, was considered. Following a simple analysis, it was considered that if the value decreased the respective period corresponded to the emission of a V symbol, and the contrary for an H. Using this methodology, we calculated the prediction accuracy for each power trace,
\begin{equation}
     \text{PA}(\%) = \frac{\text{Nº of Correctly Guessed Symbols}}{\text{Total Nº of Symbols in the Key}}\times100,
    \label{eq:Prediction_Accuracy}
\end{equation}
where a $100\%$ accuracy corresponds to the guessing of an HV sequence at the corresponding repetition frequency. We note that, during the characterization of the SoC's power consumption, it was observed that the CPU writing operation, detailed in \cite{Stanco2022VersatileSystems}, influenced the power consumption. This periodic operation caused a destabilization of the power consumption, therefore, to calculate the prediction accuracy, symbols emitted during this period were not considered. An example of the effect this operation had on the traces can be seen in Fig. \ref{fig:10kPowerTrace}, at around $3.12$ ms.
\par
\begin{figure}[ht]
    \centering
    \includegraphics[width=0.8\linewidth]{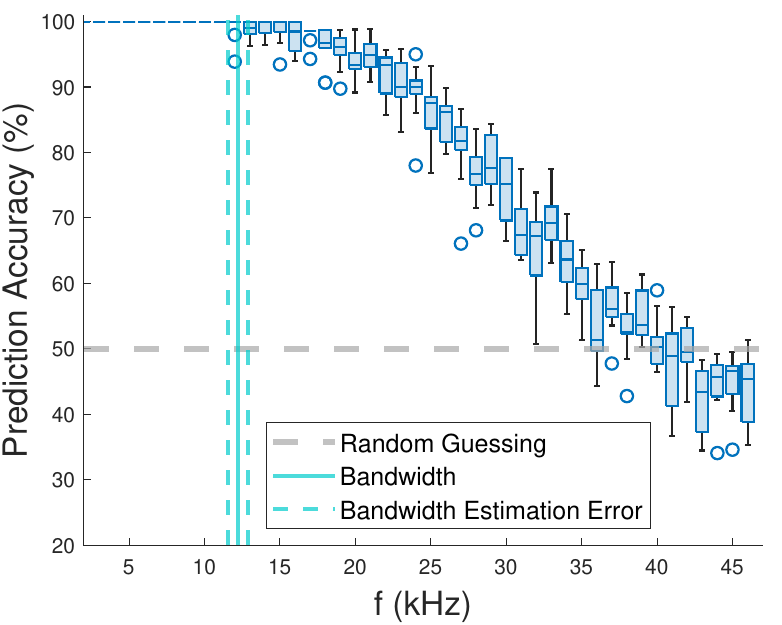}
    \caption{Box plot displaying prediction accuracy across different power traces and repetition frequencies. The system's bandwidth is presented along with its $95$\% confidence interval.}
    \label{fig:PredictionAccuracies}
\end{figure}

Fig \ref{fig:PredictionAccuracies} depicts a box chart with the prediction accuracy distribution for sequences taken at different repetition frequencies. For each frequency, 10 power traces were analyzed, and for each trace, a prediction accuracy calculated. For $f_{\text{delay}} \leq 11$ kHz we obtained a prediction accuracy of $100\%$ for all the power traces. For $f_{\text{delay}} \geq 12$ kHz, we observed a slow decrease in the prediction accuracy, which is in agreement with the expected behaviour for frequencies above the calculated bandwidth of the power system, where there is a cutoff in the distinguishability between H and V emission. 

%-----------------------------------------------------------------------------------------------Frequency Spectrum Analysis-----------------------------------------------------------------------------------------------%
\section{Frequency Spectrum Analysis of the Power Consumption}
\label{sec:FrequencySpectrum}
In this section, the frequency spectrum of the power traces is characterized in order to identify possible information leakages. Given that the focus is on the frequency domain, the bandwidth of the acquisition system is of crucial importance. Therefore, the power traces were acquired with the two SIGLENT SP$2035$A voltage probes with a $350$ MHz bandwidth, using the setup schematized in Fig \ref{fig:expsetup_osciprobes}. Nonetheless, data taken with the same setup, but with two Tektronix P$2220$ voltage probes with a bandwidth of $200$ MHz, was also considered. Although having a lower bandwidth, these probes can be impedance matched to the oscilloscope, which can only switch between an input impedance of $50$ $\Omega$ or $1$ M$\Omega$. This, together with the fact that these probes do not have an in-built signal attenuation, unlike the $350$ MHz probes, reduces the noise associated with the data acquisition. Data taken with these two probes will be analyzed and compared. Unless stated otherwise, the measurements were taken with the $350$ MHz probes.

\subsection{Fixed Sequences}
\begin{figure}[ht]
    \centering
    \includegraphics[width=0.8\linewidth]{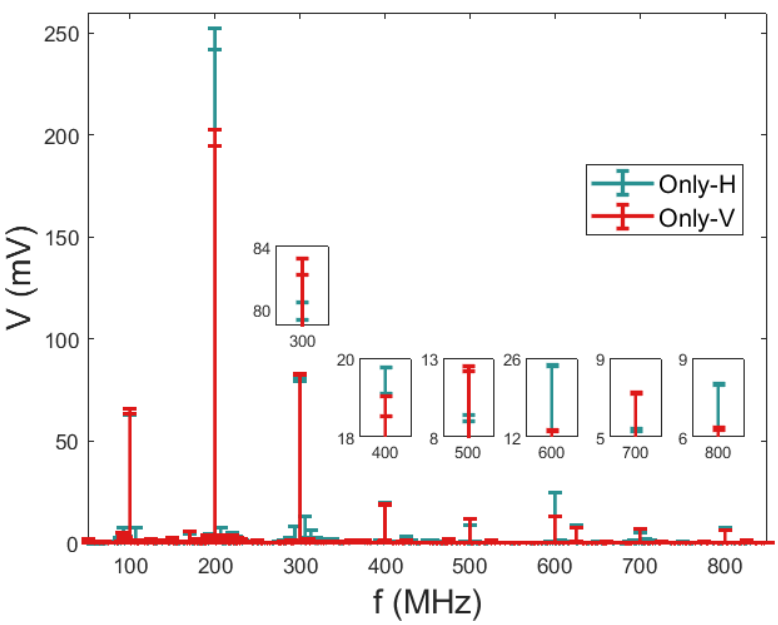}
    \caption{Average spectrum for the emission of Only-H and Only-V sequences in the $[50,900]$ MHz frequency range.}
    \label{fig:FFT_AVG_OnlyH_OnlyV}
\end{figure}

The average frequency spectrum of an Only-H and an Only-V sequence is shown in Fig. \ref{fig:FFT_AVG_OnlyH_OnlyV}, averaged over the Fast Fourier Transforms (FFTs) of $50$ power traces.
Each frequency bin displays the average magnitude, together with its standard deviation. We note that the magnitudes for all the frequencies in this range which are integer multiples of $100$ MHz show complete distinguishability between Only-H and Only-V. 
\par
\begin{figure}[ht]
    \centering
    \includegraphics[width=0.8\linewidth]{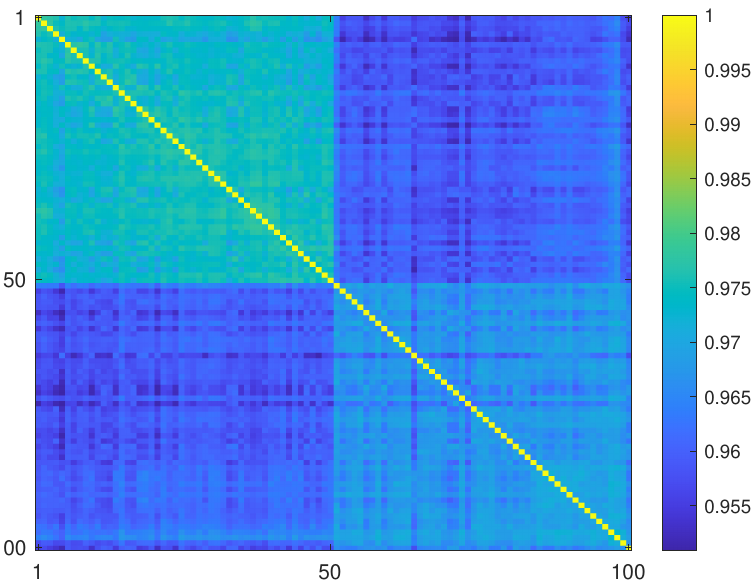}
    \caption{Correlation matrix for the power consumption's FFTs during the emission of $200000$ symbols taken from an Only-H and an Only-V emission. Only the $[50,900]$ MHz range was considered.}
    \label{fig:corr_FFT_partial_OnlyH_OnlyV}
\end{figure}

Averaging the FFTs corresponding to emissions of the same sequence directly results in a smoothing of the noise in the spectrum of the sequence. The correlation coefficients, at zero delay, between all the possible pairs of FFTs can give insight into how this noise affects the distinguishability between spectra. The coefficients were organized in matrix form such that
\begin{equation}
\left[
\begin{array}{ccc|ccc}
    h_1,h_1 & \cdots & h_1,h_{50} & h_1,v_1 & \cdots & h_1,v_{50} \\
    \vdots & \ddots & \vdots &  \vdots & \ddots & \vdots \\
    h_{50},h_1 & \cdots & h_{50},h_{50} & h_{50},v_1 & \cdots & h_{50},v_{50} \\ \hline 
    v_1,h_1 & \cdots & v_1,h_{50} & v_1,v_1 & \cdots & v_1,v_{50} \\
   \vdots & \ddots & \vdots &  \vdots & \ddots & \vdots \\
    v_{50},h_1 & \cdots & v_{50},h_{50} & v_{50},v_1 & \cdots & v_{50},v_{50}
\end{array}
\right],
\end{equation}
where $x_i$ represents the FFT of the $i^{th}$ power trace taken during Only-$X$ emission, $X \in \{\text{H},\text{V}\}$, and $x_i,x_j$ the correlation value between $x_i$ and $x_j$. The results considering the $[50,900]$ MHz frequency range are displayed in Fig. \ref{fig:corr_FFT_partial_OnlyH_OnlyV}. The top-left and bottom-right $50\times50$ sub matrices exhibit higher values, therefore showcasing that it is possible to distinguish between the spectra of Only-H and Only-V sequences without noise filtering.

The correlation coefficients were also computed considering the full frequency range spectra. In this case, all FFTs showcased a higher degree of similarity, thus it can be that the main information leakage lies in the $[50,900]$ MHz spectral range.
\par
In a realistic symbol-to-symbol side-channel attack, where the transmitted key is a random string of symbols, FFTs of smaller size sequences must be considered. The power consumption's frequency spectrum during the emission of a large quantity of symbols can only transmit information if all the symbols are either the same, as in the previous case, or differ in a known way, e.g. Only-HV sequence. Nonetheless, as the size of the sequence decreases, the FFT resolution will worsen, where
\begin{equation}
    \text{FFT Resolution} = \frac{f_{\text{sample}}}{\text{Record Length}},
\end{equation}
and $f_{\text{sample}} = 2.5$ GSa/s.
\par
We also note that the average spectra obtained for a $n$ symbol emission extracted from a fixed sequence does not necessarily correspond to the one obtained for the same $n$ symbol emission during a random sequence exchange. In fact, possible memory effects during the exchange of fixed sequences may influence the spectra of the symbols emitted. Therefore, the next section focuses on characterizing the power consumption's spectrum during the exchange of random qubits, i.e a realistic QKD scenario

\subsection{Sequences In Random Sequence Emission}
The study of different sequences' spectra during random emission was done using a data set consisting of $1$ million randomly emitted symbols and the power consumption of the SoC during their emission. For this, $5$ power traces were collected, each taken during the emission of an independent random raw key consisting of $200000$ symbols. After data acquisition, the five power traces were normalized. The spectra of different sequences were studied considering the following methodology:
\begin{enumerate}
    \item Consider sequences $S_L^i$, where $L$ is the number of symbols in the sequence and $i$ the number of possible sequences of size $L$, $i \in \{1,\ldots,2^L\}$.
    \item Divide the symbols in the data set into the $2^L$ possible sequences.  
    \item Divide the power consumption data accordingly.
    \item Calculate the FFTs of the occurrences of each sequence and average them to produce the sequence's FFT average $F_L^i$. 
    \item Compute the correlation matrix of the FFT averages
    \begin{equation}
        \begin{bmatrix}
            F_L^1,F_L^1 & \cdots  & F_L^1,F_L^{2^L}\\
            \vdots & \ddots &  \vdots\\
            F_L^{2^L},F_L^1 & \cdots &  F_L^{2^L},F_L^{2^L}
        \end{bmatrix}   .
    \end{equation}
\end{enumerate}
\par
\begin{figure}[ht]
    \centering
    \includegraphics[width=0.8\linewidth]{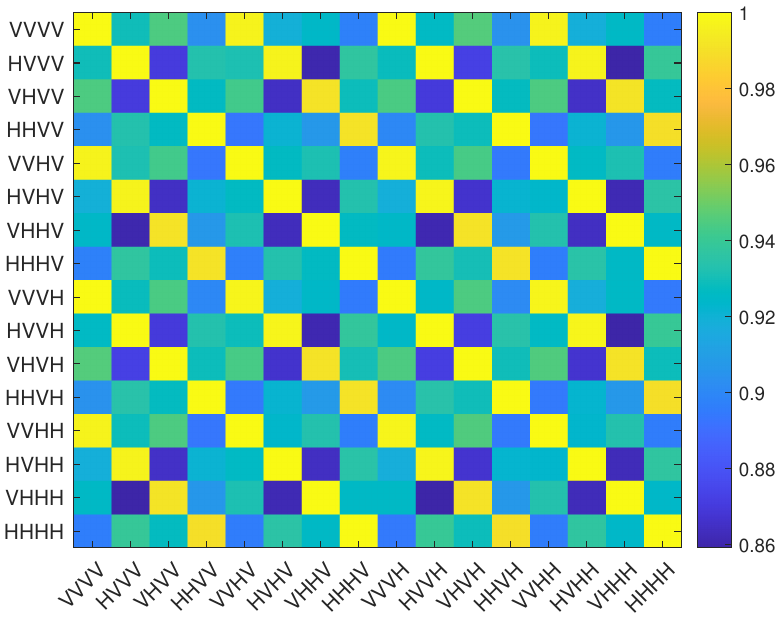}
    \caption{Correlation matrix for the FFT averages of $4$ symbol sequences, considering average phase values.}
    \label{fig:phase_L=4_CorrMatrices}
\end{figure}

By averaging all the FFTs corresponding to the emission of the same sequence, two outcomes are achieved. Firstly, the noise is smoothed out, increasing distinguishability. Secondly, since all occurrences of a given sequence were preceded and succeeded by other completely random sequences, any possible memory effect is eliminated from the spectrum.
\begin{figure}[ht]
    \centering
    \includegraphics[width=0.8\linewidth]{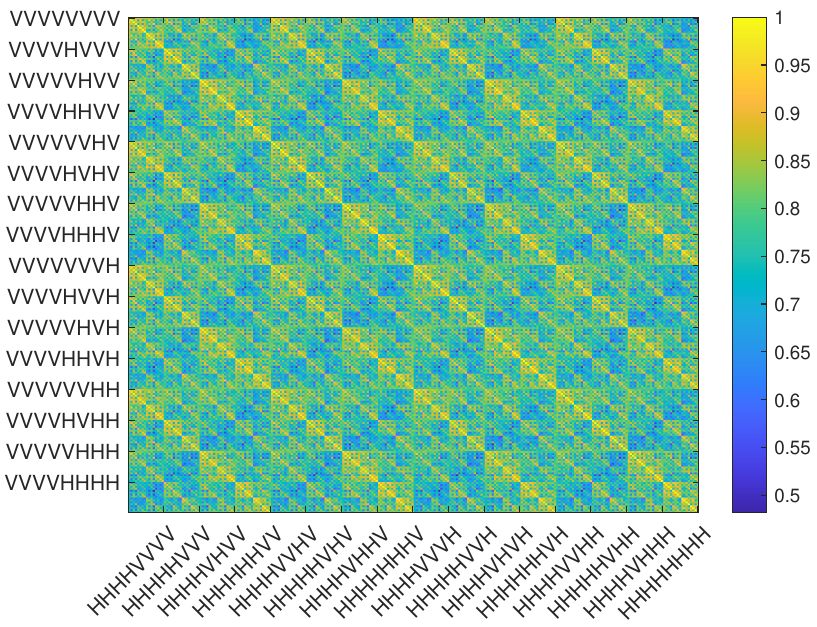}
    \caption{Correlation matrix for the FFT averages of $8$ symbol sequences, considering average phase values.}
    \label{fig:phase_L=8_CorrMatrices}
\end{figure}
Correlation matrices were computed for for $L = \{2,4,6,8\}$ and for all the considered values of $L$, the same trend was found: correlations between the frequency spectra of two sequences were higher the more initial symbols the two sequences shared. It was also found that this distinguishability was increased by considering the phase values at each frequency. In this case, the phases of each sequence occurrence were calculated, and then averaged by considering all the occurrences of the same sequence. Fig. \ref{fig:phase_L=8_CorrMatrices} and Fig. \ref{fig:phase_L=4_CorrMatrices} show the correlation matrices considering the average phases for $L=8$ and $L=4$, respectively. 
\par
To address the correlation matrices, it is useful to define the $k$-diagonal of a matrix $A$ as the set of entries $a_{i,j}$ for which $j = i + k$. In the correlation matrices, the $\pm(2^n\cdot k)$-diagonals showcase the higher correlations between the average phases of FFTs corresponding to sequences which share $n$ initial symbols, with $n<L$ and $k = 2i+1$, $i\in\{0,...,L-n-1\}$. For example, in the $L=4$ case, the $\pm 4,\pm 12$-diagonals showcase the correlation values between the average phases of FFTs corresponding to sequences which share the first $2$ symbols. In fact, by applying a threshold to the correlation matrix, it is observed that the values in these diagonals are smaller that the ones in the $\pm 8$-diagonals, which represent correlations between sequences which share the first three initial symbols. 
\par
The relationship between the average shape of a sequence's spectrum and its initial values results in an information leakage that can be exploited at high repetition rates. By comparing the spectrum of unknown sequences of size $L$ to the known average spectra of all possible sequences of size $L$, it is possible to infer information about the initial values of the unknown sequence. This concept forms the basis of the strategy implemented in the next section.

%-----------------------------------------------------------------------------------------------Transmitted Qubits at 100 MHz-----------------------------------------------------------------------------------------------%
\section{Discovering the Transmitted Qubits at $100$~MHz}
\label{sec:Discovering the Transmitted Qubits}
In this section, the information leakage found in the frequency spectrum of the power consumption of the SoC is exploited in order to discover the transmitted qubits at a repetition rate of $100$ MHz. 
\par
We propose a template side-channel attack to the SoC based on the 'FFT fingerprints' of the device. Here, we assume that the eavesdropper has initial access and control of the QKD transmitter, or of an identical copy, and uses it to store power consumption data and the corresponding sequences. With the power consumption data, the eavesdropper computes the FFT fingerprints, where for an arbitrary $L$, we define FFT fingerprint as the average FFT of the power consumption during the emission of a certain sequence with $L$ number of symbols. Note that, for our attack, we considered the average phase at each frequency bin, as opposed to the magnitude. 
\par
In the second stage of the attack, Eve loses control over the QKD transmitter, but keeps monitoring the power consumption of the FPGA. It is in this stage that Alice and Bob use the QKD system to create shared raw keys, not knowing that Eve has had previous access and is still monitoring the power consumption of the FPGA. When qubit transmission starts, Eve computes the FFTs of power consumption data corresponding to the emission of $L$ symbols. To guess these symbols, she uses the FFT fingerprints, and calculates which is more similar to the FFT of the symbols she does not know. She then uses this probability to guess part of the symbols, and performs this over the power consumption data taken during the emission of the raw key she wants to guess. 
\par
In the proposed side-channel attack strategy there are two parameters which Eve must choose: $L$, the length of the sequences for which the FFT fingerprints are calculated, and $\delta N$, the number of symbols she guesses after finding the best match for an FFT, such that $0< \delta N\leq L$.
\par
To prove the feasibility of this side-channel attack, we implemented it to discover $15$ independent random raw keys. The previously obtained set of $1$ million random symbol emissions was used to calculate the FFT fingerprints. Each raw key we wished to discover consisted of $200000$ random symbols and to quantify the performance of our attack we calculated the prediction accuracy, defined in Eq. \ref{eq:Prediction_Accuracy}. The prediction accuracy with a random guessing method, this is, for each symbol in an infinite sized raw key, one randomly predicts between H and V, would be $50\%$. Therefore, we consider the discovering of a raw key successful if it exceeds random guessing by $3$ standard deviations of the binomial distribution \cite{Baliuka2023Deep-learning-basedDistribution}, this is,
\begin{equation}
    \text{Prediction Accuracy} (\%) >  50.34\%.
    \label{eq:sucessful_hack}
\end{equation}
\par
\begin{figure}[ht]
    \centering
    \includegraphics[width=0.9\linewidth]{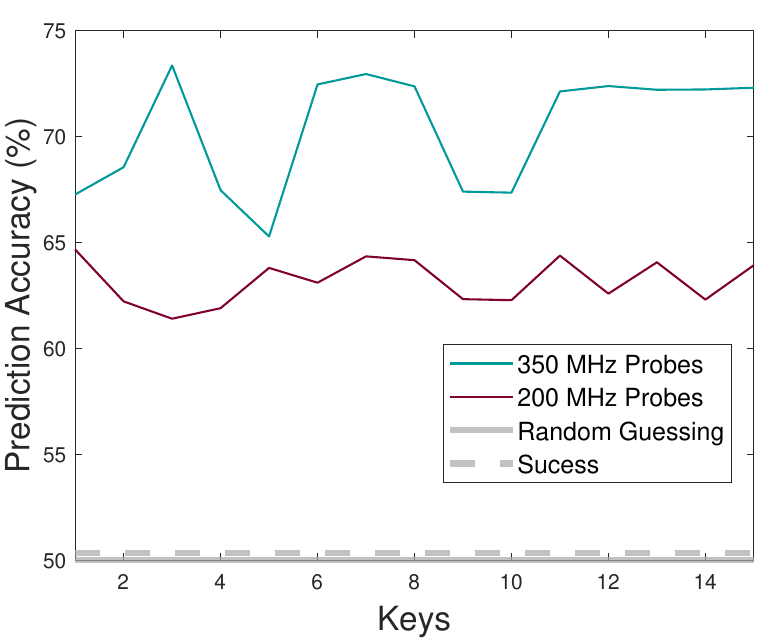}
    \caption{Prediction Accuracy for a $L=8$ and $\delta N =1$ strategy.}
    \label{fig:PA_L8_N1}
\end{figure}

Two side-channel attack trials were performed, one with a pair of oscilloscope probes with a bandwidth of $350$ MHz, and another with a probe pair with $200$ MHz bandwidth. The trials were spaced by a month and, for each, a different set of $15$ independent random raw keys was considered. The FFT fingerprints were calculated using a data set acquired with the corresponding probes. With a strategy with $L=8$ and $\delta N = 1$, we were able to achieve a maximum prediction accuracy of $73.35\%$. The prediction accuracy achieved for each of the $15$ sequences in the two trials is shown in Fig \ref{fig:PA_L8_N1}. The performance of the trial with the $350$ MHz probes surpasses the performance of the one with the $200$ MHz probes, which tells us that the bandwidth of the acquisition system is more important to the side-channel attack performance than the amount of noise associated with the measurements. 

Side-channel attack strategies with $L=\{2,4,6\}$ and $\delta N =1$ were also performed to the same data. In all of them we see a decrease in the prediction accuracy compared to the $L=8$, $\delta N = 1$ case, although all lied above the success threshold. Generally, for fixed $\delta N$, we expected the performance to decrease as $L$ decreases, given that the FFT resolution worsens. However, although this is true for $L=\{8,4,2\}$, the $L=6$ strategy performed worse than the $L=4$ one. The former entails $\text{FFT Resolution} = 16.67$ MHz, meaning that the FFTs computed do not have frequency bins located at the integer multiple frequencies of the $100$ MHz frequency. For $L=4$, these frequency bins exist, which might justify the better performance of a strategy with this value. Additionally, for $L=4$ and $L=2$, the side-channel attack performed better for the data acquired with the $200$ MHz probes, meaning that, as the spectral resolution decreases, noise becomes a larger impediment for guessing the key and the bandwidth of the system a less important requisite. 
\par
\begin{figure}[ht]
    \centering
    \includegraphics[width=0.9\linewidth]{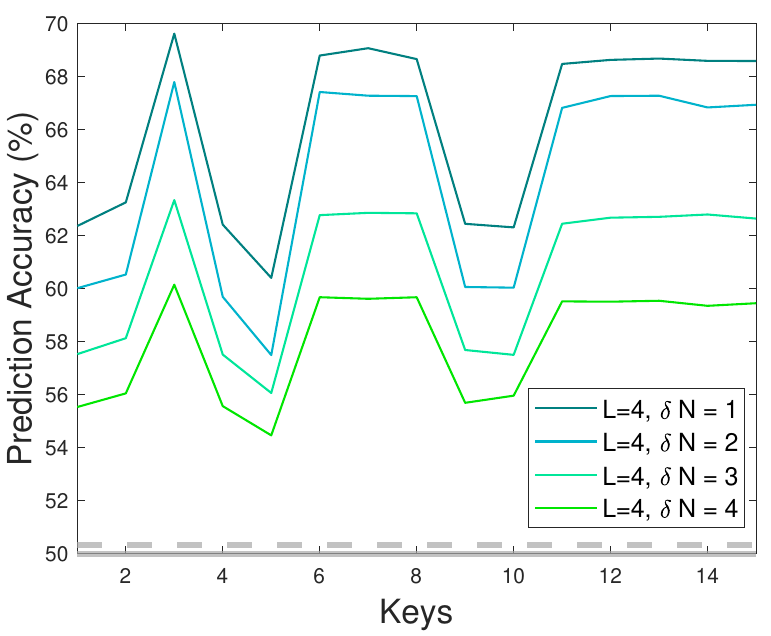}
    \caption{Prediction Accuracies for side-channel attack strategies with different $\delta N$ and constant $L=4$.}
    \label{fig:PA_L=4_VaryN}
\end{figure}

To study the impact of $\delta N$, strategies with fixed $L$ and different values of $\delta N$ were considered. The results for $L=4$ are shown in Fig. \ref{fig:PA_L=4_VaryN}, where for simplicity, only the trial with the $350$ MHz probes was considered. Here, we see a decrease in the side-channel attack performance as $\delta N$ increases. This means that after calculating the FFT of a given snippet and finding its highest correlated FFT fingerprint, we have a high degree of certainty on the first symbol in the snippet. This certainty decreases for the second symbol, and so forth. Therefore, the strategy with the lowest performance was the one that combined the lowest $L$ with the highest $\delta N$ allowed, this is, $L=2$ and $\delta N = 2$. Nonetheless, even in this case, all 15 raw keys yielded a prediction accuracy above the success threshold.

%--------------------------------------------------------------------------------------------------------Conclusions------------------------------------------------------------------------------------------------------%
\section{Conclusions and Outlook}
\label{sec:Conclusions}
In this work, we have demonstrated the feasibility of a power side-channel attack to a QKD transmitter, namely by exploiting the information leakage of the SoC controlling the electro-optical components of a specific QKD transmitter. 
\par
The experimental results suggest that the average power consumption of the SoC depends on the transmitted qubit values, allowing a simple side channel attack to reach $100$\% prediction accuracy for repetition frequencies below the bandwidth associated with the current rise time.
\par
Regarding the frequency spectrum of the SoC's power consumption, for random sequences of length $L$, the spectra became increasingly similar as the number of shared initial symbols between two sequences increased. This allowed for the implementation of a template side-channel attack based on the SoC's FFT fingerprints, which was implemented to predict two sets of 15 transmitted keys emitted at $f_{\text{rep}} = 100$ MHz. The maximum prediction accuracy achieved for this case was $73.35 \%$.
\par
Notably, the side-channel attack proposed in this work requires an eavesdropper capable of measuring the power consumption of a device integrated in Alice's setup. Clearly, this would not be straightforward in a practical deployment, as it would require integrating a power-sensing setup into Alice's system. However, the information leakage found, together with the success of the side-channel attack, shows that power side-channel analysis to a SoC is a feasible technique to retrieve information about the transmitted qubits. Therefore, power consumption data from the electronics in the parties' setups must also be treated as sensitive information, making the findings in this work an important step in understanding which protective measures must be implemented to Alice's and Bob's electronics in order to guarantee the overall security of the QKD system.
\par

%--------------------------------------------------------------------------------------------------------Acknowledgments------------------------------------------------------------------------------------------------------%
\section{Acknowledgements}
The authors thank the support from project QSNP – Quantum Secure Networks Partnership (GA 101114043) of the Horizon Europe Programme of the European Commission. BC, YO and RC thank the support from FCT -- Funda\c{c}\~{a}o para a Ci\^{e}ncia e a Tecnologia (Portugal), namely through project UIDB/04540/2020 and UIDB/50021/2020.

Author MRB acknowledges support from the European Union’s Horizon Europe Framework Programme under the Marie Sklodowska Curie Grant No. 101072637, Project Quantum-Safe Internet (QSI).

The authors would like to thank F. Rampazzo for lending us the KEYSIGHT N2791A differential probe.

This work is (partially) supported by ICSC – Centro Nazionale di Ricerca in High Performance Computing, Big Data and Quantum Computing, funded by European Union – NextGenerationEU

\bibliography{references}% Produces the bibliography via BibTeX.

\end{document}